\documentclass[
    10pt,
    prd,
    amsmath,
    nofootinbib,
    amssymb,
    aps,
    showkeys,
]{revtex4-2}
\usepackage{fix-cm}
\usepackage[utf8]{inputenc}
\usepackage[english]{babel}
\usepackage{graphicx}
\usepackage{mathrsfs}
\usepackage{booktabs}
\usepackage{geometry}
\usepackage{latexsym,amsthm,amsfonts}
\usepackage{makeidx}   
\usepackage[T1]{fontenc}
\usepackage{subfigure}
\usepackage{subcaption}
\usepackage{dcolumn}
\usepackage{bm}
\usepackage{cancel}
\usepackage{xcolor}
\usepackage[hyperindex,colorlinks,linkcolor=blue,citecolor=red]{hyperref}

\usepackage{textgreek}
\usepackage{comment}

\allowdisplaybreaks[1]

\usepackage{upgreek}
\usepackage{graphicx}
\usepackage{tensor}
\usepackage{soul}
\usepackage[normalem]{ulem}

\begin{document}
\title{Light deflection and gravitational lensing effects inspired by loop quantum gravity}

	\author{A. R. Soares}
    \email{adriano.soares@ifma.edu.br}
	\affiliation{Grupo de Estudos e Pesquisas em Laborat\'orio de Educa\c{c}\~ao matem\'atica, Instituto Federal de Educa\c{c}\~ao Ci\^encia e Tecnologia do Maranh\~ao,  R. Dep. Gast\~ao Vieira, 1000, CEP 65393-000 Buriticupu, MA, Brazil.}

\author{C. F. S. Pereira}
	\email{carlosfisica32@gmail.com}
	\affiliation{Departamento de Física, Universidade Federal do Espírito Santo, Av. Fernando Ferrari, 514, Goiabeiras, 29060-900, Vit\'oria, ES, Brazil.}
		
	\author{R. L. L. Vit\'oria}
    \email{ricardovitoria@professor.uema.br/ricardo-luis91@hotmail.com}
	\affiliation{Departamento de Ci\^encias Exatas e Naturais, Universidade Estadual do Maranh\~ao, Contorno da Av. Jo\~ao Alberto de Sousa, Ramal, 65700-000, Bacabal, MA, Brazil.}
    \affiliation{Centro de Ci\^encias Humanas, Naturais, Sa\' ude e Tecnologia, Universidade Federal do Maranh\~ao, Estrada Pinheiro/Pacas, Enseada, 65200-000, Pinheiro, MA, Brazil.}

\author{Marcos V. de S. Silva}
	\email{marcosvinicius@fisica.ufc.br}
	\affiliation{Departamento de Física, Programa de Pós-Graduação em Física,
Universidade Federal do Ceará, Campus Pici, 60440-900, Fortaleza, Ceará, Brazil}

\author{H. Belich} 
\email{humberto.belich@ufes.br}
\affiliation{Departamento de F\'isica e Qu\'imica, Universidade Federal do Esp\'irito Santo, Av.Fernando Ferrari, 514, Goiabeiras, Vit\'oria, ES 29060-900, Brazil.}

\begin{abstract}
\noindent  
In the present work, we theoretically investigate light deflection in the weak and strong field regimes for two regular spacetimes with corrections from loop quantum gravity. We treat analytically the expansions for both limits and use them as a basis for investigating gravitational lensing observables. We analyze and provide reasonable values for observables related to the second model that observational tools may be able to detect.
\end{abstract}
	
\keywords{}
	
\maketitle
	
\section{Introduction}\label{sec1}

It is well known in the literature that Einstein's general relativity has a variety of models that present problems of geodesic singularities, thus motivating the search for theories of gravity that have more general characteristics \cite{INTRO1,INTRO2,INTRO3,INTRO4,INTRO5,INTRO6,INTRO7,INTRO8,INTRO9}. If we analyze only from the perspective of models that have as their purpose the removal of singularities inside the black hole (BH), we have a variety of solutions for regular BHs in the context of general relativity that emerged around 1968 with James Bardeen \cite{INTRO10} and a wide variety of other models have been proposed over the years \cite{INTRO11, INTRO12, INTRO13, INTRO14, INTRO15, INTRO16, INTRO17, INTRO18, INTRO19, INTRO20, INTRO21, INTRO22, INTRO23, INTRO24, INTRO25}.
In addition to the regular BH solutions mentioned above, a new class of regular solutions, called Black-Bounce, was recently proposed by Simpson and Visser \cite{SV}, and depending on the relationship between the model parameters, some possible configurations can be established: $a>2m$ two-way traversable wormhole (WH), $a=2m$ a one-way traversable WH with the throat located at the origin $r=0$, and if $a<2m$ we have a WH hidden by two symmetric horizons. In this sense, this new class of regular solutions has been explored in the most varied scenarios \cite{INTRO26,INTRO27,INTRO28,INTRO29,INTRO30,INTRO31,INTRO31-1,INTRO32,INTRO33,INTRO34,INTRO35,INTRO36,INTRO37,INTRO38,INTRO38-2,INTRO39,INTRO40,INTRO41,INTRO42,INTRO43,INTRO43-2}.

On the other hand, in the context of an extended gravitational theory, loop quantum gravity (LQG) emerges as an alternative proposal to introduce the process of singularity removal via the polymerization mechanism. Even though it is not a perturbative theory, LQG is not a complete theory from the point of view of spacetime quantization in the vicinity of singularities and therefore has been explored in the low-energy scenario for effective models \cite{INTRO44,INTRO45,INTRO46,INTRO47,INTRO48,ADAILTON,EXT1,EXT2,JCAP1,JCAP2}. Recently, geodesically complete regular BH solutions have been explored for anomaly-free holonomy corrections \cite{INTRO49,INTRO50,INTRO51,INTRO52,INTRO53,INTRO54,EXT3,EXT4,EXT5,EXT6}. Furthermore, in \cite{BASE} the authors study the greybody factors and the evaporation of a series of Black-Bounce spacetimes, such as the Simpson-Visser (SV) solution \cite{SV} and others inspired by LQG, with the aim of addressing the dark matter problem through LQG corrections.

In this work, we investigate the theoretical aspects of gravitational lensing effects in two regular spacetimes incorporating LQG corrections. These spacetimes can also represent a WH, and we analyze the deflection of light in the strong and weak field regimes for both solutions. Regarding the regularized D’Ambrosio-Rovelli (DR) model, we analyze gravitational lensing effects for the strong and weak field regimes, in addition to providing plausible values for the observables.

The gravitational lensing process can be seen as an effect resulting from the interaction mechanism between light and some gravitational field, causing the light beam to have its trajectory altered \cite{INTRO55, INTRO56}. This effect can be minimal when the light beam passes very far from the source of matter (weak field regime), but the deflection can also be intensified when the light beam passes very close to the source, reaching a limit region in which the deflection angle diverges (strong field regime). In the 1960s, there were some pioneering works \cite{INTRO57, INTRO58}, which showed that relativistic images were very weak, in addition to not having expressions that adequately described the strong field regime. It was not until the 2000s that Virbhadra and Ellis \cite{KS1} developed reasonably consistent gravitational lensing equations for the strong field regime. Later, Bozza presented a methodology for studying lensing effects in the strong field regime \cite{BOZZA}, which was improved by Tsukamoto \cite{JAPA1}. Since then, the phenomenology of gravitational lenses has been explored in a variety of contexts, such as BHs \cite{INTRO59,INTRO60,INTRO61,INTRO62,INTRO63,INTRO64,INTRO65,INTRO66,INTRO67,INTRO68,EXT7,EXT8,JCAP3,JCAP4,JCAP5,JCAP6,JCAP7,JCAP8}, WHs \cite{INTRO69,INTRO70,INTRO71,INTRO72,INTRO73,INTRO74,INTRO75,INTRO76,INTRO77,INTRO78,INTRO79}, topological defects \cite{INTRO80,INTRO81,INTRO82,INTRO83,INTRO84}, modified gravity \cite{INTRO85,INTRO86,INTRO87,INTRO88,INTRO89,INTRO90,INTRO91,INTRO92,INTRO93}, regular BHs, and Black-Bounces \cite{INTRO94,INTRO95,INTRO96,INTRO97,INTRO98,INTRO99,INTRO100}.

In Section \ref{sec2} of the present work, we present the general relations for a spherically symmetric metric and conserved quantities. In Sections \ref{sec3} and \ref{sec4}, we introduce two models and analyze the deflection of light in the strong and weak field regimes. In Section \ref{sec5}, we apply the gravitational lensing technique to the second model and analyze some observables. In Section \ref{sec6}, we present the final considerations and conclusions.

\section{General relations}\label{sec2}

We will take as a starting point a spherically symmetric and static line element given by
\begin{equation}\label{1}
ds^2= -f(r)dt^2 + \frac{dr^2}{f(r)} +\Sigma^2(r)\left(d\theta^2+\sin^2\theta{d\phi^2}\right),
\end{equation} where $f\left(r\right)$ and $\Sigma\left(r\right)$ are functions that depend only on the radial coordinates. Thus, we define a smooth curve in this generic spacetime, Eq. (\ref{1}), that has length $S$ which is given by
\begin{equation}\label{2}
S= \int \sqrt{\left(g_{\mu\nu}\frac{dx^\mu}{d\lambda}\frac{dx^\nu}{d\lambda}\right)}d\lambda,
\end{equation} where $\lambda$ is an affine parameter that can represent the observer's proper time. Taking $S$ as the affine parameter itself, we can show that the curve that minimizes Eq. (\ref{2}) is the same one that minimizes
\begin{equation}\label{3}
\int \left(g_{\mu\nu}\frac{dx^\mu}{d\lambda}\frac{dx^\nu}{d\lambda}\right)d\lambda=\int\mathcal{L}{d\lambda}.
\end{equation}

Therefore, considering the analysis in the equatorial plane, $\theta=\frac{\pi}{2}$, the Lagrangian becomes:

\begin{equation}\label{4}
\mathcal{L}=-f(r)\left(\frac{dt}{d\lambda}\right)^2 + \frac{1}{f(r)}\left(\frac{dr}{d\lambda}\right)^2 + \Sigma^2(r)\left(\frac{d\phi}{d\lambda}\right)^2.
\end{equation}

Applying the Euler-Lagrange equations to the above expression, we define the quantities conserved at time $t$ and at $\phi$. Therefore, we have to
\begin{equation}\label{5}
L= \Sigma^2(r)\frac{d\phi}{d\lambda},  \qquad \qquad E=f(r)\frac{dt}{d\lambda}.
\end{equation}

Thus, substituting the conserved quantities, Eq. (\ref{5}), in Eq. (\ref{4}) and considering only null geodesics, $\mathcal{L}=0$, leads (\ref{4}) to

\begin{equation}\label{6}
\left(\frac{dr}{d\lambda}\right)^2= E^2- \frac{L^2{f(r)}}{\Sigma^2(r)}.
\end{equation}

The above expression can be compared with the dynamics of a classical particle with energy $E$ and subjected to an effective potential $V_{eff}=\frac{L^2f(r)}{\Sigma^2}$ that depends on the metric functions. In general, we can obtain the radius of the photon sphere through $\frac{dV_{eff}}{dr}=0$.

\section{FIRST MODEL}\label{sec3}

We will develop the methodology of light deflection in a large gravity system, considering as background the Peltola-Kunstatter (PK) spacetime \cite{BASE} inspired by LQG. The PK spacetime regularizes the BH interior using a semi-classical polymerization scheme, in which the momentum operators are replaced by holonomic variables. In the context of dilatonic gravity, the energy term in the Hamiltonian is modified, and the resulting change in the dynamics of the dilaton field leads to a bounce instead of a classical geodesic singularity inside the BH. Thus, we have that the metric functions that form this line element taking as a basis the general metric Eq. (\ref{1}) are $f(r)= \frac{r-2M}{\sqrt{r^2+a^2}}$ and $\Sigma^2(r)=r^2+a^2$. This geometry has an internal structure similar to a WH that has an event horizon located at $r_H=2M$ and its throat at $r=0$. In some way, we can say that this throat $a$ is hidden by the presence of the horizon. Thus, the radius of the photon sphere for this model is given by
\begin{equation}\label{7}
r_{m}= \frac{3M}{2} \pm \frac{\sqrt{9M^2+2a^2}}{2}.
\end{equation}

Note in Eq. (\ref{7}) that we have two photon spheres, however, as we are interested in studying lensing only in the region outside the BH, which corresponds to $r_m>r_H$. We will use the radius $r_{m2}$ only as an auxiliary way to expand the parameter $\Lambda_1$\footnote{The parameter $\Lambda_1$ will be defined later.} in the strong field regime. It is important to note that $r_{m1}$ corresponds to a maximum point in the effective potential, which means that the photon orbits there are unstable, while $r_{m2}$ corresponds to a minimum point in the effective potential, thus representing a stable photon orbit. However, this stable orbit is hidden by the event horizon. Therefore, the radius of the photon sphere is given by 
\begin{equation}\label{8}
    r_{m1}= \frac{3M}{2} + \frac{\sqrt{9M^2+2a^2}}{2}, 
\end{equation}

\begin{equation}\label{9}
     r_{m2}= \frac{3M}{2} - \frac{\sqrt{9M^2+2a^2}}{2}.
\end{equation}
Note that in the limit where we consider the bounce radius tending to zero, $a\to{0}$, the radius Eq. (\ref{8}) becomes the radius of the photon sphere for the Schwarzschild BH (SBH), $r_{m1}=3M$. This ingredient will be extremely useful in analyzing the deviation of light in the weak field regime (distant observer) and the strong field regime.

\subsection{Expansion for light  deflection in the weak field limit}\label{sec31}

We will then consider a photon that begins its trajectory in an asymptotically flat region and approaches the BH at a distance $r_0$ from the center of the BH, called the turning point, such that $r_0 > r_{m1}$. When deflected by the presence of this gravitational field of the BH, the photon tends to return to another asymptotically flat region. At the turning point, we have $V_{eff}(r_0)= E^2$, which leads to the following expression
\begin{eqnarray}\label{10}
\frac{1}{\beta^2}= \frac{r_0- 2M}{\left(r^2_0 +a^2\right)^{3/2}},
\end{eqnarray} where $\beta=\frac{L}{E}$ represents the impact parameter. Thus, the orbit equation Eq. (\ref{6}) can be written as
\begin{eqnarray}\label{11}
\frac{dr}{d\phi}= \pm\frac{\Sigma^2(r)}{L}\left[E^2-\frac{L^2f(r)}{\Sigma^2(r)}\right]^\frac{1}{2}.
\end{eqnarray}

Considering that the angular contributions in $\phi$ before and after the turning point are the same, we have the following
\begin{eqnarray}\label{12}
\Delta\phi=\pm 2\int^{\infty}_{r_0}dr\left[\frac{\Sigma^4(r)}{\beta^2}-\Sigma^2(r)f(r)\right]^{-\frac{1}{2}}.
\end{eqnarray}

Let us now introduce a change of coordinates in Eq. (\ref{12}), where $u=\frac{1}{\sqrt{r^2+a^2}}$. Note that now the integration limits are defined as $u\to{0}$ for $r\to\infty$ and $u\to{u_0}=\frac{1}{\sqrt{r^2_0+a^2}}$ for $r\to{r_0}$. Thus, the equation for the angular deviation in the new coordinates is expressed by
\begin{eqnarray}\label{13}
\Delta\phi=\pm 2\int^{u_0}_0{du}\left[\left(1-a^2u^2\right)\left(\frac{1}{\beta^2}+2Mu^3-u^2\sqrt{1-a^2u^2}\right) \right]^{-\frac{1}{2}}.
\end{eqnarray}

Considering the exchange of coordinates that was performed above in the expression of the impact parameter Eq. (\ref{10}), we have $\frac{1}{\beta^2}=u^2_0\sqrt{1-a^2u^2_0}-2Mu^3_0$. Thus, we used the impact parameter in this new coordinate in Eq. (\ref{13}) and assume that in the weak field limit the photon passes very far from the BH. Then, we can consider the approximation $M\ll{1}$ and $a\ll{1}$ and expand the orbit expression Eq. (\ref{13}) in terms of up to second order in the throat parameter $a$, we see that the deviation of the light is given by $\delta\phi=\Delta\phi-\pi$:
\begin{eqnarray}\label{14}
\delta\phi \simeq \frac{4M}{\beta} +\frac{5\pi{a^2}}{8\beta^2} + \frac{Ma^2}{\beta^3}\left[\frac{29}{3}-\frac{5\pi}{4}\right].
\end{eqnarray}

For the regular spacetimes mentioned in the original article \cite{BASE}, the procedure begins with a line element that has LQG effects described in the coordinates $\bar{r}$, and then they are analyzed in another coordinate system $\bar{r}=\sqrt{r^2+a^2}$ so that the metric functions begin to acquire similarities with the SV spacetime. Thus, in the deviation of the light to the weak field regime described in the equation above, we can observe that the radius of the WH throat contributes to this effect. Thus, in the limit in which we have no contributions from the LQG, $a\to{0}$, the deviation to the SBH is recovered.

\subsection{Deflection of light in the strong field limit}\label{sec32}

To develop the gravitational deflection of light in the strong field regime, we will use the calculation methodology developed by Bozza \cite{BOZZA} and extended by Tsukamoto \cite{JAPA1}. To do this, we begin by considering the following exchange of variables $z=1-\frac{r_0}{r}$ in the orbit equation Eq. (\ref{12}). So, we can rewrite it as
\begin{eqnarray}\label{15}
\Delta\phi(r_0)=\pm\int^{1}_{0}\frac{2r_0{dz}}{\sqrt{G(z,r_0)}},
\end{eqnarray}
where,
\begin{eqnarray}\label{16}
G(z,r_0)= \frac{r^4_0}{\beta^2} +\frac{a^4(1-z)^4}{\beta^2}+\frac{2a^2r^2_0(1-z)^2}{\beta^2}-\left[r_0-2M(1-z)\right]\sqrt{r^2_0 +a^2(1-z)^2}(1-z)^2.
\end{eqnarray}

Performing the power series expansion for the function $G(z,r_0)$ at the point where $z=0$, which is equivalent to saying that $r\to{r_0}$ and considering only terms up to second order $\mathcal{O}(z^2)$, we have that
\begin{eqnarray}\label{17}
G(z,r_0)\simeq \Lambda_1(r_0){z} +\Lambda_2(r_0){z^2},
\end{eqnarray} where the expansion parameters $\Lambda_1$ and $\Lambda_2$ are defined as
\begin{eqnarray}\label{18}
\Lambda_1(r_0)&=& \frac{2r_0\left(r^2_0 -3Mr_0 -\frac{a^2}{2}\right)}{\sqrt{r^2_0 +a^2}}= \frac{2r_0{(r_0-r_{m2})(r_0-r_{m1})}}{\sqrt{r^2_0 +a^2}},\\\label{19}
\Lambda_2(r_0)&=& \frac{6a^4r_0 + 30Ma^2r^2_0 -5a^2r^3_0 + 12Mr^4_0 -2r^5_0}{2\left(r^2_0 + a^2\right)^{3/2}}.
\end{eqnarray}

In the strong field regime, we have the limit where $r_0\to{r_{m1}}$ and therefore the expansion coefficients, Eqs. (\ref{18},\ref{19}), in this limit, are expressed by:
\begin{eqnarray}\label{20}
\Lambda_1(r_0) \to \Lambda_1(r_{m1})&=&0,\\
\Lambda_2(r_0) \to \Lambda_2(r_{m1})&=&\frac{\gamma^2\left[54M^4\left(3M+\sqrt{9M^2+2a^2}\right) + 18M^2a^2\left(4M+\sqrt{9M^2+2a^2}\right) \right]}{\sqrt{6}\left(-M+\sqrt{9M^2+2a^2}\right)}+ \nonumber \\\label{21}
&+&\frac{\gamma^2\left[a^4\left(8M+\sqrt{9M^2+2a^2}\right) \right]}{\sqrt{6}\left(-M+\sqrt{9M^2+2a^2}\right)},
\end{eqnarray} in which $\gamma^2=\frac{-M+\sqrt{9M^2+2a^2}}{\left(3M^2+a^2+M\sqrt{9M^2+2a^2}\right)^{3/2}}$.

The expansion coefficients, Eqs. (\ref{20},\ref{21}), in the strong field regime, $r \to r_{m1}$, cause the integral in Eq. (\ref{15}) to exhibit a logarithmic divergence and therefore it is necessary to separate this integral into two parts: a regular part $\Delta\phi_R(r_0)$ and a divergent part $\Delta\phi_D(r_0)$. Therefore, we have to
\begin{eqnarray}\label{22}
\Delta\phi(r_0)=\Delta\phi_R(r_0)+\Delta\phi_D(r_0).
\end{eqnarray}

Thus, the divergent part is defined by
\begin{eqnarray}
\Delta\phi_D(r_0)&=& \int^1_0\frac{2r_0}{\sqrt{\Lambda_1(r_0)z+\Lambda_2(r_0)z^2}}= \nonumber \\\label{23}
&-&\frac{4r_0}{\sqrt{\Lambda_2(r_0)}}\log\left(\sqrt{\Lambda_1(r_0)}\right) + \frac{4r_0}{\sqrt{\Lambda_2(r_0)}}\log\left(\sqrt{\Lambda_2(r_0)}+ \sqrt{\Lambda_1(r_0)+\Lambda_2(r_0)}\right).
\end{eqnarray}

To have some control over the divergent part, we need to expand the coefficient $\Lambda_1(r_0)$ and the impact parameter $\beta(r_0)$ in the radius of the photon sphere $r_{m1}$. Thus, Eqs. (\ref{10}) and (\ref{18}) become
\begin{eqnarray}
\beta(r_0) &\simeq & \frac{3^{3/4}}{2^{1/4}\gamma} + \frac{2^{3/4}}{3^{1/4}\gamma}\nonumber \\
&\times& \left[\frac{\left(2Ma^2+9M^3+(3M^2+a^2)\sqrt{9M^2+2a^2}\right)}{\left(-M+\sqrt{9M^2+2a^2}\right)\left(3M^2+a^2+M\sqrt{9M^2+2a^2}\right)^2}\right]\left(r_0-\frac{3M}{2}-\frac{\sqrt{9M^2+2a^2}}{2}\right)^2, \nonumber \\\label{24}
\end{eqnarray} note in the equation of the impact parameter above, Eq. (\ref{24}), that we can rewrite as a function of the expansion parameter $\Lambda_1$ (\ref{19}) and vice versa
\begin{eqnarray}
\Lambda_1(r_0)&\simeq&\left(\frac{2^{1/4}\gamma\beta}{3^{3/4}}-1\right)^{1/2}  \nonumber \\
&\times &\left[\frac{(9M^2+2a^2)(3M+\sqrt{9M^2+2a^2})^{2}(-M+\sqrt{9M^2+2a^2})(3M^2+a^2+M\sqrt{9M^2+2a^2})}{\left(2Ma^2 + 9M^3 + (3M^2+a^2)\sqrt{9M^2+2a^2}\right)}\right]^{1/2}. \nonumber \\\label{25}
\end{eqnarray}

Thus, substituting Eqs. (\ref{21}) and (\ref{25}) in Eq. (\ref{23}) and considering the strong field regime $r_0\to{r_{m1}}$, we find
\begin{eqnarray}
\Delta\phi_D= &-& \frac{\frac{\left(3M+\sqrt{9M^2+2a^2}\right)}{2}\log\left(\frac{2^{1/4}\gamma\beta}{3^{3/4}}-1\right)}{\sqrt{\Lambda_2(r_{m1})}} +\frac{\left(3M+\sqrt{9M^2+2a^2}\right)}{\sqrt{\Lambda_2(r_{m1})}}\nonumber \\
&\times& \log\left[\frac{4\Lambda_2(r_{m1})}{\left[\frac{(3M+\sqrt{9M^2+2a^2})^2(9M^2+2a^2)(-M+\sqrt{9M^2+2a^2})(3M^2+a^2+M\sqrt{9M^2+2a^2})}{\left(2Ma^2+9M^3+(3M^2+a^2)\sqrt{9M^2+2a^2}\right)}\right]^{1/2}}\right]. \nonumber \\\label{26}
\end{eqnarray}

The equation above represents an analytical expression that refers to the divergent part of the deviation of light in the PK spacetime subjected to the effects of LQG. Logically, in the limit in which the bounce parameter tends to zero, the results of the Schwarzschild spacetime are recovered.

Regarding the regular part, Eq. (\ref{22}), we have to perform the analysis of the limit at which the impact parameter Eq. (\ref{10}) becomes critical $r_0\to{r_{m1}}$ and then use the expressions Eq. (\ref{16}) and Eq. (\ref{19}) also in this limit. Thus, the general expression of the regular part becomes
\begin{eqnarray}\label{27}
\Delta\phi_R= \int^{1}_0 \frac{2r_{m1}}{\sqrt{G(z,r_{m1})}}dz -\int^{1}_{0}\frac{2r_{m1}}{\sqrt{\Lambda_2(r_{m1})}}\frac{dz}{z}.
\end{eqnarray}

\begin{figure}[htb!]
\centering
	{\includegraphics[width=0.7\textwidth]{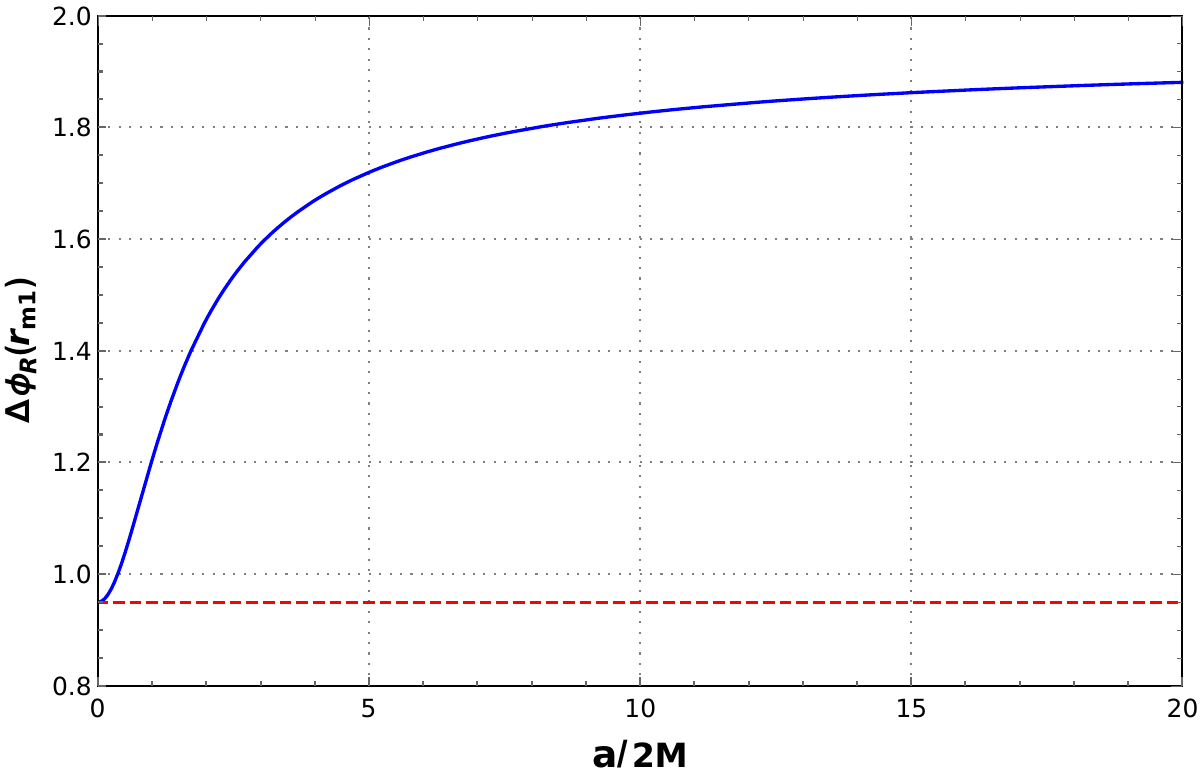}}
    \caption{Representation of the regular part of the integration when $\beta=\beta_c$.}
\label{RMOD1}
\end{figure}

The expression for the regular part, Eq. (\ref{27}), has no analytical solution and therefore requires a numerical evaluation. Note in Fig. (\ref{RMOD1}) that the blue curve represents the behavior of the regular part of the angular deviation for the LQG effects and the dotted red curve represents the regular part for the Schwarzschild limit $a=0$. In this limit, the regular part of the integration tends to converge to $\approx{0.9496}$. 

\begin{figure}[htb!]
\centering  
	\subfigure[]{\label{campoSV1}
	{\includegraphics[width=0.47\linewidth]{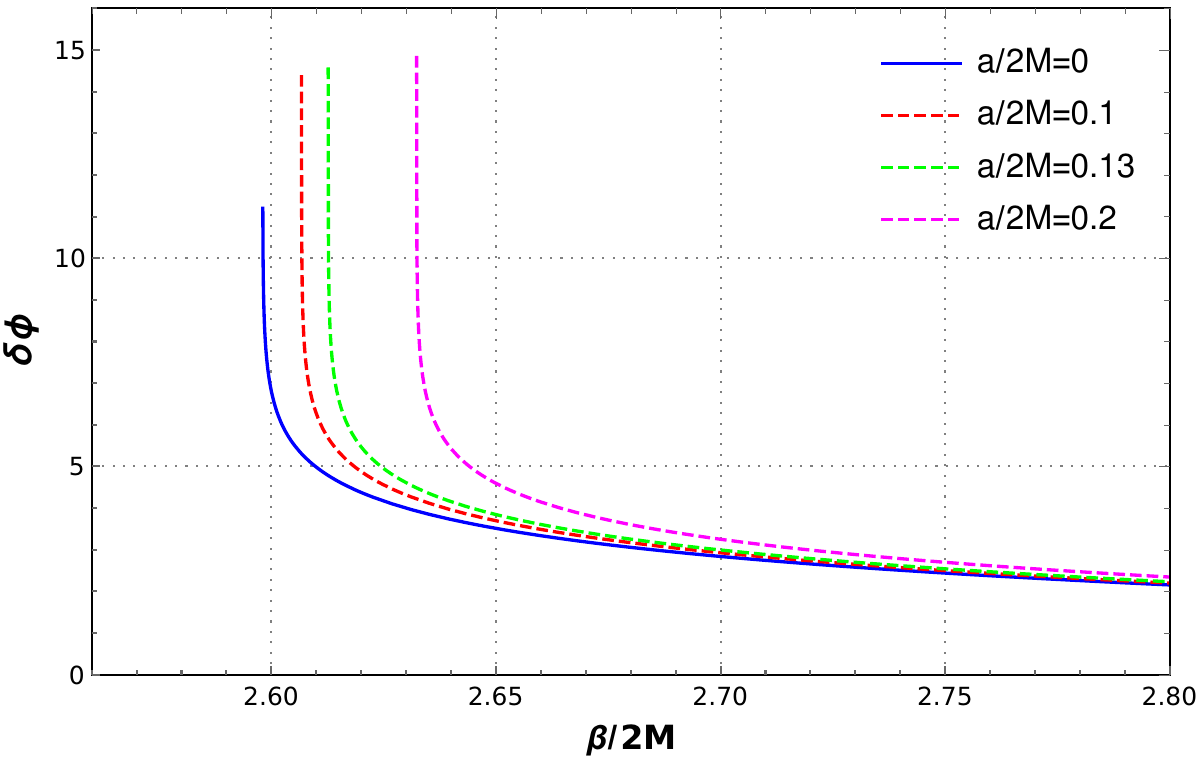}}}\qquad
	\subfigure[]{\label{poteSV1}
	{\includegraphics[width=0.47\linewidth]{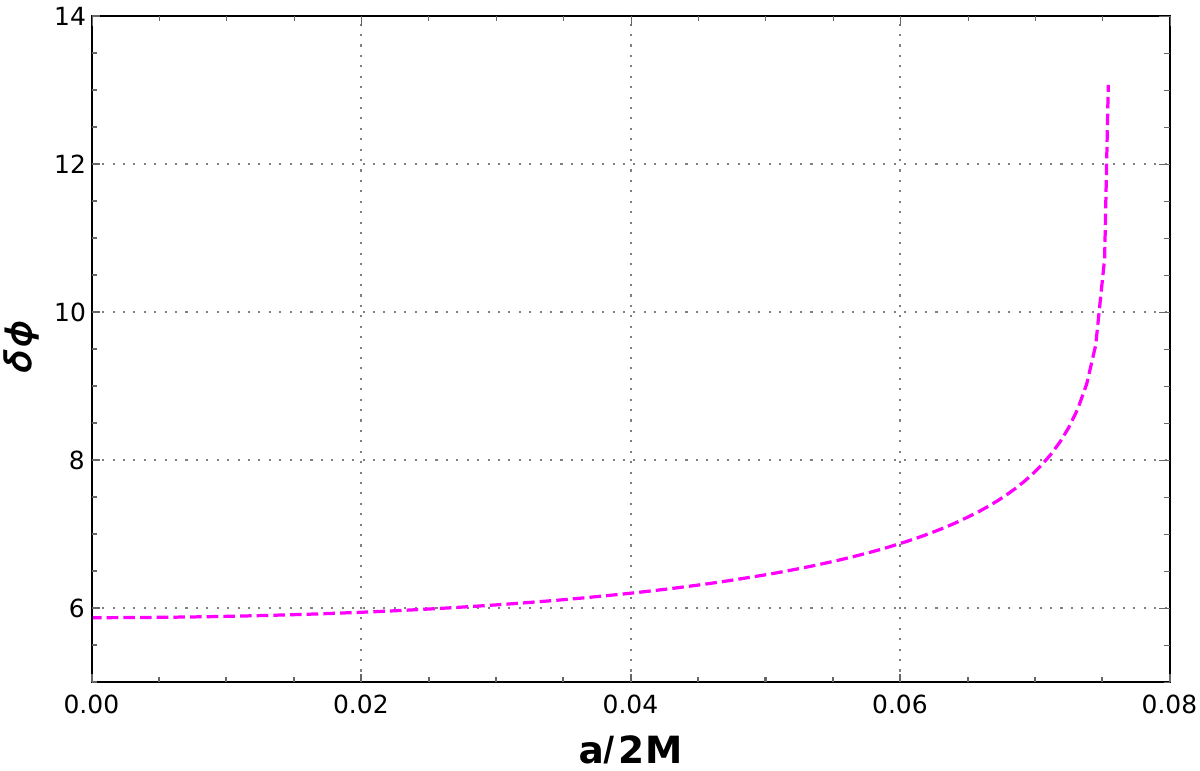}}}
\caption{In (a) we have the deflection angle as a function of the impact parameter for some value of the throat radius \textbf{a}. In (b) the deflection angle as a function of $a/2M$ for $\beta/2M=\frac{3\sqrt{3}}{2}+0.005$.}
\label{DESVIOMOD1}
\end{figure}

Thus, the deviation of light in the PK spacetime modified by the effects of LQG \cite{BASE} is defined by $\delta\phi=\Delta\phi-\pi$ and, since there is no analytical form for the regular part of Eq. (\ref{27}), a numerical analysis is necessary. In the left panel in Fig. (\ref{DESVIOMOD1}), we have the graphical representation of the angular deviation as a function of the impact parameter, considering some values for the relationship between the bounce parameter $a$ and the BH mass $M$. In the limit of the bounce parameter tending to zero, the SBH results are recovered. In the right panel of Fig. (\ref{DESVIOMOD1}), we have the deviation of light, for $\beta/2M=\frac{3\sqrt{3}}{2}+0.005$, as a function of $a/2M$. We can observe divergences in $\delta \phi$. These divergences occur due to the proximity between the particles' impact parameter and the critical impact parameter.

\section{SECOND MODEL}\label{sec4}

In \cite{BASE}, this spacetime was investigated in the context of LQG, in which authors, through changes in the dynamic variables of the metric, show how the gravitational field evolves naturally through the central singularity in a solution that is the time reversal of the BH interior, namely a white hole interior. This solution is particularly interesting because it sheds light on the problem of the BH information paradox. Therefore, consider the line element below as the DR spacetime with LQG corrections
\begin{eqnarray}\label{DB1}
ds^2= - \left(1-\frac{2M}{\sqrt{r^2+a^2}}\right)dt^2 +  \left(1-\frac{2M}{\sqrt{r^2+a^2}}\right)^{-1}\left(1+\frac{\bar{a}}{\sqrt{r^2+a^2}}\right)dr^2 + \Sigma^2(r)d\Omega^2, 
\end{eqnarray} where $\Sigma^2(r)=r^2+a^2$ and $d\Omega^2$ is the same area element defined in (\ref{sec2}). We emphasize that the parameters $\bar{a}$ and $\textbf{a}$ are equal. We label the parameter as $\bar{a}$ only as a way of verifying the consistency of the calculations so that in the limit in which it tends to zero, it is necessary to recover the results for the SV spacetime, as is the case of the calculation of light deviation and lensing effects investigated in Ref. \cite{SOARES1}. In the original article \cite{BASE}, such a distinction does not exist, and therefore, if we turn off the LQG effect, the geometry is reduced to the SBH.

The conserved quantities, angular momentum and energy of the system, for this model are defined in the same way as in Eq. (\ref{5}). Thus, the equation regarding the radial direction of a light beam is given by:
\begin{eqnarray}\label{DB2}
\left(\frac{dr}{d\lambda}\right)^2= \left(1+\frac{\bar{a}}{\sqrt{r^2+a^2}}\right)^{-1}\left[E^2-\frac{L^2}{(r^2+a^2)}\left(1-\frac{2M}{\sqrt{r^2+a^2}}\right)\right].
\end{eqnarray} 

As in the first model, we can compare the expression above with the classical dynamics of a particle subjected to an effective potential. Thus, the radius of the photon sphere of this spacetime is the same as that associated with the SV spacetime that was discussed in Ref. \cite{SOARES1}. Therefore, the radius of the photon sphere is given by

\begin{eqnarray}\label{DB3}
r_m= \pm\sqrt{(3M)^2-a^2}.
\end{eqnarray}

The structure of this regularized DR spacetime is the same as that described in the original paper by Simpson and Visser \cite{SV}. It depends only on the relationship between the parameters that describe the model. If $a>2M$ we have a bidirectional WH, $a=2M$ a unidirectional WH, and if $a<2M$ we have the case of a regular BH with symmetric horizons and a throat located at the origin $r=0$. Therefore, in the limit where the throat of this spacetime tends to zero, the radius of the SBH photon sphere, $r_m=3M$, is recovered.

\subsection{Expansion for light deflection in the weak field limit}\label{sec41}

As in the first model, to begin the analyses in the weak field regime, we then consider a light beam moving from an asymptotically flat region toward the center of the BH called the turning point, such that $r_0> r_m$. When deflected by the presence of the gravitational field, the photon tends to return to another asymptotically flat region. Therefore, at the point of return we have $V_{eff}(r_0)=E^2$ and the relationship between energy and angular momentum is given by
\begin{eqnarray}\label{DB4}
\frac{1}{\beta^2}= \frac{\left(\sqrt{r^2_0 +a^2}-2M\right)}{\left(r^2_0 + a^2\right)^{3/2}}.
\end{eqnarray}

Thus, using the same exchange of coordinates established just before, Eq. (\ref{12}), we have that the expression referring to the angular deviation, Eq. (\ref{DB2}), becomes
\begin{eqnarray}\label{DB5}
\Delta\phi = \pm 2\int^{u_0}_0{du}\left[\frac{(1-a^2u^2)}{(1+\bar{a}u)}\left(\frac{1}{\beta^2}-u^2(1-2Mu)\right)\right]^{-\frac{1}{2}}.
\end{eqnarray}

For the coordinate exchange considered above, the impact parameter, Eq. (\ref{DB4}), is defined by $\frac{1}{\beta^2}=u^2_0(1-2Mu_0)$. Thus, substituting the impact parameter into the equation of angular deviation of this BH (\ref{DB5}) and considering that in the weak field limit, the approximations $M\ll{1}$ and $a\ll{1}$ are valid. Therefore, performing the expansion of equation (\ref{DB5}) and considering only terms up to second order in the bounce, we find that the deviation of the light in the weak field regime is defined by $\delta\phi=\Delta\phi-\pi$, thus
\begin{equation}
\delta\phi \simeq \frac{2(2M+\bar{a})}{\beta} + \frac{(12\pi{M\bar{a}} + 2\pi{a^2}-16M\bar{a})}{8\beta^2}  
+ \frac{a^2\left(16M+4\bar{a}-3\pi{M}\right)}{6\beta^3} +\frac{M\bar{a}a^2\left(15\pi-32\right)}{16\beta^4}.\label{DB6}
\end{equation}

The expression above represents the deviation of light for a BH in the weak field regime being influenced by the effects of LQG \cite{BASE}. As mentioned above, the parameter $\bar{a}$ was introduced in the line element Eq. (\ref{DB1}) just so that it would be possible to perform the comparison of the calculation regarding the deviation of light in the SV spacetime for the limit of $\bar{a}\to{0}$ \cite{SOARES1}. Thus, if we consider in Eq. (\ref{DB6}) the limit of $\bar{a}\to{0}$, we recover the same gravitational deflection for the light studied in the SV spacetime in \cite{SOARES1}
\begin{eqnarray}\label{DB7}
\delta\phi_{SV} \simeq \frac{4M}{\beta} + \frac{\pi{a^2}}{4\beta^2} + \frac{Ma^2(16-3\pi)}{6\beta^3}.
\end{eqnarray}

In a complementary way, if we consider the line element Eq. (\ref{DB1}) in the format in which it is addressed with the LQG correction \cite{BASE}, we must consider the limit at which $\bar{a}\to{a}$. Thus, the expression for the light deviation, Eq. (\ref{DB6}), becomes
\begin{eqnarray}\label{DB8}
\delta\phi_{LQG} \simeq  \frac{2(2M+a)}{\beta} + \frac{a(6M\pi + a\pi-8M) }{4\beta^2} + \frac{a^2(16M+4a-3M\pi)}{6\beta^3}.
\end{eqnarray}

To verify the consistency of the results, we can consider the limit at which the LQG effects cease to exist $a\to{0}$. Therefore, the deviation of the light towards the SBH is recovered. In the present work, we are considering only orbits for the photon in which $a<3M$. However, there is the possibility of a more complete analysis in future works considering the regions $a>3M$ and $a=3M$, as performed in Ref. \cite{N+}.

\subsection{Deflection of light in the strong field limit}\label{sec42}

To describe the light beam in the strong field regime, we start by substituting the conserved quantities Eq. (\ref{5}) in Eq. (\ref{DB2}) and using the following coordinate exchange $z=1-\frac{r_0}{r}$. Then, we have
\begin{eqnarray}\label{DB9}
\Delta\phi(r_0)=\pm\int^{1}_{0}\frac{2r_0{dz}}{\sqrt{G(z,r_0)}},
\end{eqnarray}
where,
\begin{eqnarray}
G(z,r_0)= \frac{\left[r^2_0 +a^2(1-z)^2\right]^{5/2}}{\left[\sqrt{r^2_0+a^2(1-z)^2}+\bar{a}(1-z)\right]}\left[\frac{1}{\beta^2}-\frac{(1-z)^2\left(\sqrt{r^2_0 +a^2(1-z)^2}-2M(1-z)\right)}{\left[r^2_0 +a^2(1-z)^2\right]^{3/2}}\right]. \nonumber \\\label{DB10}
\end{eqnarray} Note that, as a consistency check of the results, we can take the limit $\bar{a} \to 0$, in which case we recover the result for the SV spacetime \cite{SOARES1}.

Performing the power series expansion of the function Eq. (\ref{DB10}) around the origin, and considering only terms up to second order in the variable $\mathcal{O}(z^2)$, we have that
\begin{eqnarray}\label{DB11}
G(z,r_0) \simeq \Lambda_1(r_0)z +  \Lambda_2(r_0)z^2,
\end{eqnarray} 
where the expansion parameters are defined by
\begin{eqnarray}
\Lambda_1(r_0)&=& \frac{2r^2_0\left(r^2_0 + a^2 -3M\sqrt{r^2_0 + a^2}\right)}{\left(r^2_0 + a^2 + \bar{a}\sqrt{r^2_0 +a^2}\right)},\label{DB12} \\
\Lambda_2(r_0)&=&  \frac{r^2_0\sqrt{r^2_0 + a^2}\left[a^2(15M\bar{a}-6r^2_0)-r^4_0 -5a^4\right] + r^2_0(r^2_0 + a^2)\left[r^2_0(6M+\bar{a})+5a^2(3M-\bar{a})\right]}{(r^2_0 + a^2)^{3/2}\left(\bar{a}+\sqrt{r^2_0 + a^2}\right)^2}. \nonumber \\\label{DB13}
\end{eqnarray}

Considering the expansion coefficients, Eqs. (\ref{DB12}) and (\ref{DB13}), in the limit where $r_0 \to r_m$, the integral Eq. (\ref{DB9}) acquires a logarithmic divergence and therefore it is necessary to separate it into two parts, a regular one $\Delta\phi_R(r_0)$ and a divergent one $\Delta\phi_D(r_0)$, so that $\Delta\phi(r_0)= \Delta\phi_R(r_0) + \Delta\phi_D(r_0)$. Thus, the expansion coefficients, in this limit, are given by
\begin{eqnarray}\label{DB14}
\Lambda_1(r_0) &\to & \Lambda_1(r_m)=0, \\\label{DB15}
\Lambda_2(r_0) &\to & \Lambda_2(r_m)= \frac{(9M^2-a^2)^2}{3M(3M+\bar{a})}.
\end{eqnarray}

Thus, the divergent part is given by
\begin{eqnarray}
\Delta\phi_D(r_0)&=& \pm \int^1_0 \frac{2r_0}{\sqrt{\Lambda_1(r_0)z+\Lambda_2(r_0)z^2}}= \nonumber \\
&-&\frac{4r_0}{\sqrt{\Lambda_2(r_0)}}\log\left(\sqrt{\Lambda_1(r_0)}\right)  + \frac{4r_0}{\sqrt{\Lambda_2(r_0)}}\log\left(\sqrt{\Lambda_2(r_0)}+ \sqrt{\Lambda_1(r_0)+\Lambda_2(r_0)}\right).\label{DB16}
\end{eqnarray}

To have some control over the divergent part, we need to expand the coefficient $\Lambda_1(r_0)$ and the impact parameter $\beta(r_0)$ in the radius of the photon sphere $r_{m}$. Thus, Eqs. (\ref{DB4}) and (\ref{DB12}) become
\begin{eqnarray}\label{DB17}
\beta(r_0) &\simeq & 3\sqrt{3}M + \frac{\sqrt{3}}{2M}\left(\sqrt{r^2_0 + a^2}-3M\right)^2, \\\label{DB18}
\Lambda_1(r_0) & \simeq & \frac{2\sqrt{6}M(9M^2-a^2)}{3(3M+\bar{a})}\left[3 +\sqrt{6}\left(\frac{\beta}{\sqrt{27M^2}}-1\right)^{1/2}\right]\left(\frac{\beta}{\sqrt{27M^2}}-1\right)^{1/2}.
\end{eqnarray}

Therefore, substituting Eqs. (\ref{DB17}) and (\ref{DB18}) in Eq. (\ref{DB16}), in the strong field regime, $r_0 \to r_m $, we have that
\begin{eqnarray}
\Delta\phi_D= &-& \sqrt{\frac{3M(3M+\bar{a})}{(9M^2-a^2)}}\log\left(\frac{\beta}{\sqrt{27M^2}}-1\right)  \nonumber \\
&+& 2 \sqrt{\frac{3M(3M+\bar{a})}{(9M^2-a^2)}}\log\left[\frac{2(9M^2-a^2)}{\sqrt{6}M^2\left[3 + \sqrt{6}\left(\frac{\beta}{\sqrt{27M^2}}-1\right)^{1/2}\right]}\right]. \label{DB19}
\end{eqnarray}

The above expression, Eq. (\ref{DB19}), represents the divergent part of the deviation of light in spacetime for a BH under the effects of LQG. Naturally, in the limit where the parameter $\bar{a}$ tends to zero, the expression for the SV case must be recovered \cite{SOARES1}. Still considering the strong field limit, the regular part of the integration is given by
\begin{eqnarray}\label{DB20}
\Delta\phi_R= \int^{1}_0 \frac{2r_{m}}{\sqrt{G(z,r_{m})}}dz -\int^{1}_{0}\frac{2r_{m}}{\sqrt{\Lambda_2(r_{m})}}\frac{dz}{z}.
\end{eqnarray}

The contribution of the regular part above Eq. (\ref{DB20}) does not have an analytical solution and, therefore, requires a numerical evaluation. The angular deviation for the BH that is influenced by the effects of LQG is defined by $\delta\phi=\Delta\phi-\pi$ and will also be analyzed numerically.

\begin{figure}[htb!]
\centering  
	\subfigure[]{\label{}
	{\includegraphics[width=0.47\linewidth]{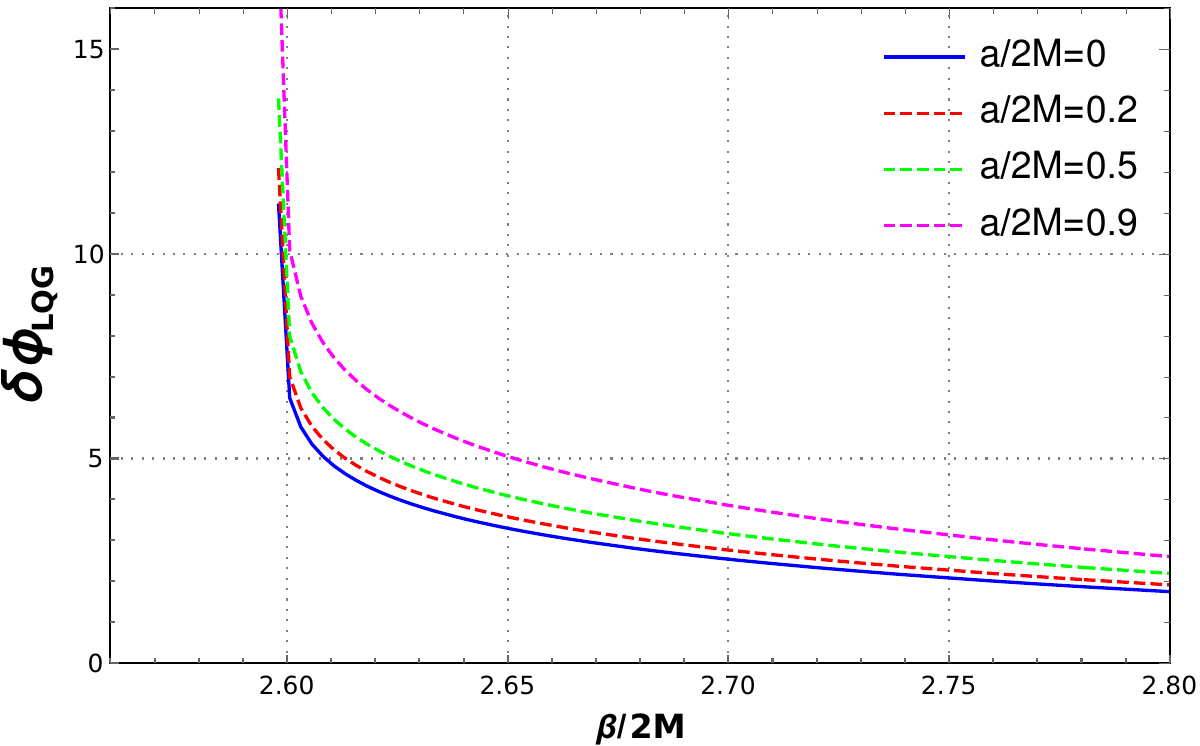}}}\qquad
	\subfigure[]{\label{}
	{\includegraphics[width=0.47\linewidth]{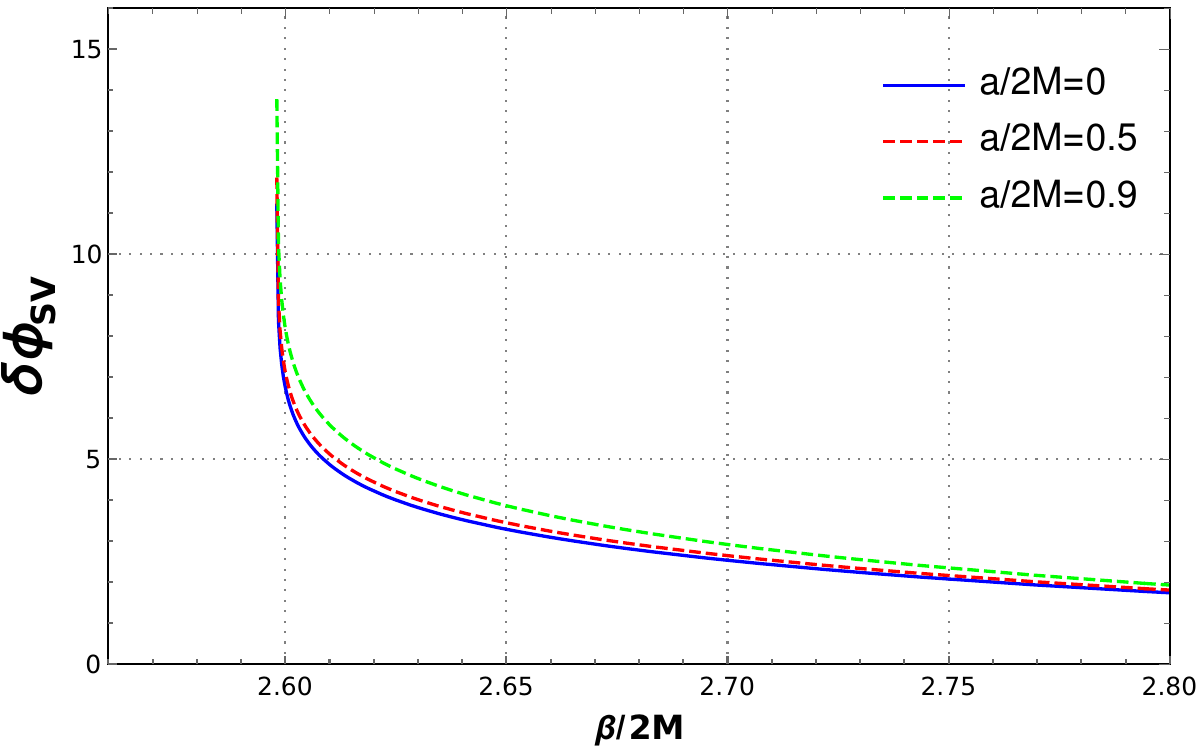}}}
\caption{Deflection angle as a function of the impact parameter for some value of the throat radius \textbf{a} in (a) DR spacetime with $\bar{a}=a$ (b) SV spacetime.}
\label{DESVIOSVMODI1}
\end{figure}

\begin{figure}[htb!]
\centering  
	\subfigure[]{\label{DBREG1}
	{\includegraphics[width=0.47\linewidth]{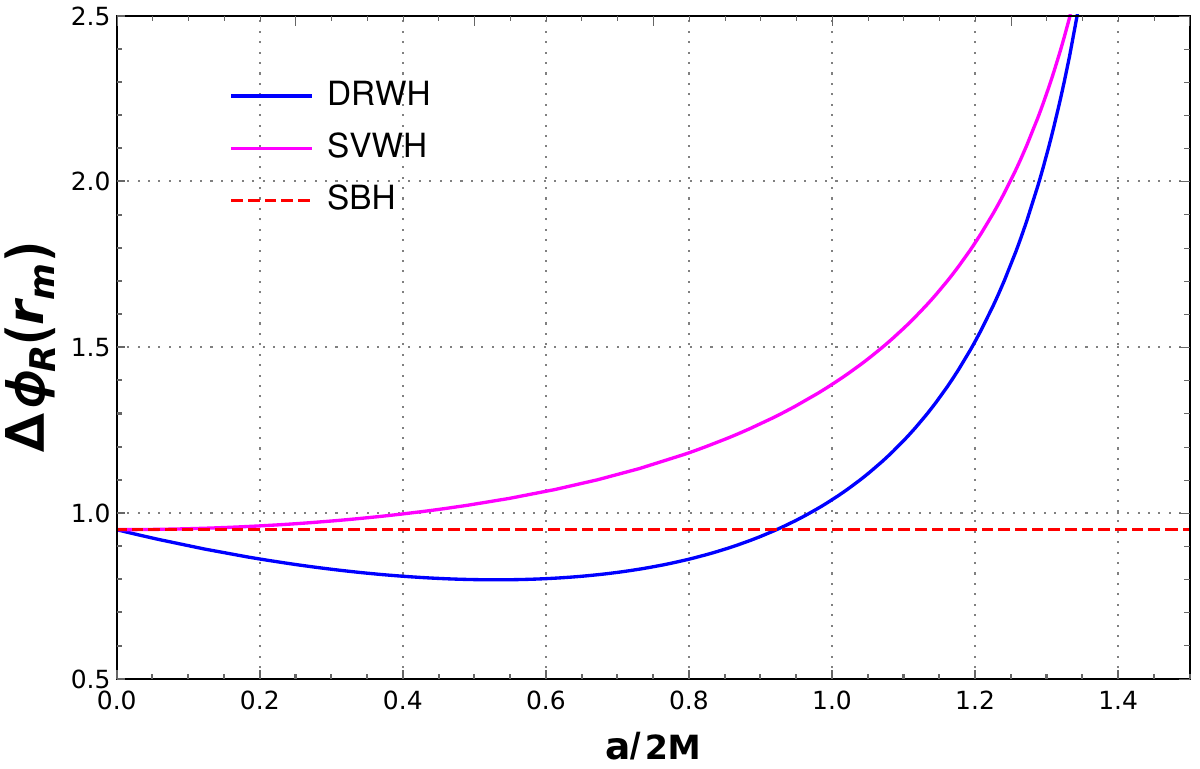}}}\qquad
	\subfigure[]{\label{DBtot1}
	{\includegraphics[width=0.47\linewidth]{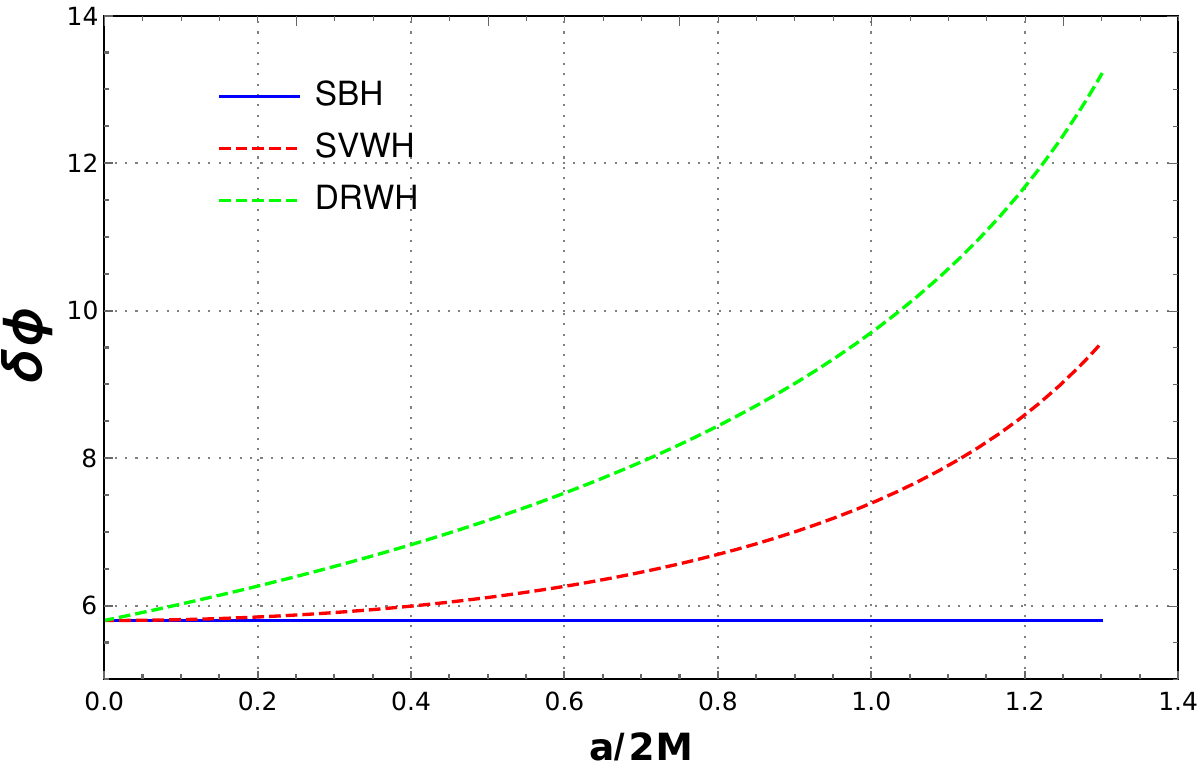}}}
\caption{In (a) we have the regular integration part. In (b) the deflection angle as a function of $a/2M$ for $\beta/2M=\frac{3\sqrt{3}}{2}+0.005$. For DB spacetime we are considering  $\bar{a}=a$.}
\label{DESVIOSVMODI2}
\end{figure}

In Fig. (\ref{DESVIOSVMODI1}) we have the angular deviation as a function of the impact parameter for some values of the radius of the WH throat. In the left panel, we have the curves referring to the LQG corrections in the DR spacetime, and in the right panel we have the behavior for the SV spacetime, in agreement with Ref. \cite{SOARES1}. The divergences arise as the chosen impact parameter approaches the critical impact parameter. In the representation of the angular deviation referring to the regular part of the integration, Fig. \ref{DBREG1}, we can see that the contribution referring to the SV space is greater than or equal to the Schwarzschild BH. Note that, in the DR case, there is one additional possibility compared to the SV case, depending on the LQG corrections, the regular part may be larger, smaller, or equal to that of SBH.

The blue curve in the left panel of Fig. (\ref{DESVIOSVMODI2}) illustrates the behavior of the regular part of the integral in Eq. (\ref{DB20}) for the DR spacetime. In the same panel, the solid magenta curve corresponds to the regular part associated with the SV spacetime \cite{SOARES1}, while the red dotted curve represents the SBH case. The right panel displays the deflection angle for these same spacetimes; however, in this case, the green curve represents the DR spacetime, the blue curve corresponds to the SBH, and the red curve represents the SV case.

It is important to note that, although in the DR spacetime model the regular contribution may be smaller than in the Schwarzschild case, Fig. \ref{DBREG1}, the physically measurable quantity is the total deflection, which is a combination of the regular and divergent parts, and this total deflection will always be equal to or greater than that of the SBH case, Fig. \ref{DBtot1}.

\section{LENS EQUATION AND OBSERVABLES}\label{sec5}

\begin{figure}[htb!]
\centering
	{\includegraphics[width=0.95\textwidth]{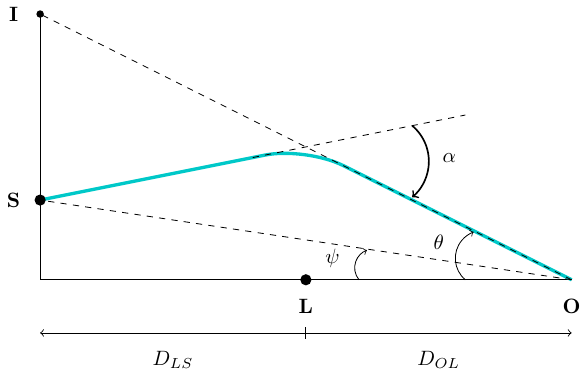}}
    \caption{Light angular deflection diagram.}
\label{LENS1}
\end{figure}

This section will be reserved to establish the connection between the study of light deflection in both the weak field regime, Eq. (\ref{DB6}), and the strong field regime, Eqs. (\ref{DB19}) and (\ref{DB20}), where $\delta\phi = \Delta\phi_D + \Delta\phi_R - \pi$, with the gravitational lensing equations for the DR spacetime. In this way, we can construct physical quantities that can, in principle, be observed. We will initially consider lensing in the strong field regime.

In Fig. (\ref{LENS1}), we have a panel visually representing the lens diagram. The light beam emitted by the source \textbf{S} is deflected by the presence of the BH located at \textbf{L} towards an observer \textbf{O}. The angular deflection of light is given by $\alpha$. The angular positions of the source and the image concerning the optical axis, $\overline{LO}$, are given, respectively, by $\psi$ and $\theta$. Thus, considering that the source \textbf{S} is practically aligned with the lens \textbf{L}, a position in which the relativistic images should be more expressive \cite{BOZZA1,KS1}, we then have the lens equation that relates the angular positions $\psi$ and $\theta$ defined as
\begin{eqnarray}\label{LT1}
\psi = \theta -\frac{D_{LS}}{D_{OS}}\Delta\alpha_n,
\end{eqnarray} where $\Delta\alpha_n$ is the deflection angle subtracted from all the loops made by the photons before reaching the observer, that is, $\Delta\alpha_n = \alpha -2n\pi$.  In this approach, we use the following approximation for the impact parameter $\beta\approx\theta{D_{OL}}$. Then the contributions to the angular deflection, Eqs. (\ref{DB19}) and (\ref{DB20}), are rewritten as:
\begin{eqnarray}\label{LT2}
\alpha(\theta)= -\bar{p}\log\left(\frac{\theta{D_{OL}}}{\beta_c}-1\right) +\bar{b},
\end{eqnarray}
where we have
\begin{eqnarray}\label{LT3}
\beta_c &=& 3\sqrt{3}M, \qquad \qquad \bar{p}=\sqrt{\frac{3M(3M+\bar{a})}{(9M^2-a^2)}}, \\\label{LT4}
\bar{b}&=& 2\bar{p}\log\left[\frac{2(9M^2-a^2)}{\sqrt{6}M^2[3+\sqrt{6}\left(\frac{\beta}{\beta_c}-1\right)^{1/2}]}\right] + \Delta\phi_R -\pi,
\end{eqnarray}
given that the regular part is given by Eq. (\ref{DB20}). 

To obtain $\Delta\alpha_n$, we expand $\alpha(\theta)$ close to $\theta = \theta^{0}_n$, where $\alpha(\theta^{0}_n)
 = 2n\pi$:
 \begin{eqnarray}\label{LT5}
\Delta\alpha_n= \frac{\partial\alpha}{\partial\theta} \Big|_{\theta=\theta^{0}_n}\left(\theta -\theta^{0}_n\right).
\end{eqnarray}

Evaluating Eq. (\ref{LT2}) on $\theta=\theta^{0}_n$, we have
\begin{eqnarray}\label{LT6}
\theta^{0}_n= \frac{\beta_c}{D_{OL}}(1+e_n), \qquad \mbox{with} \quad e_n= e^{\frac{\bar{b}-2n\pi}{\bar{p}}}.
\end{eqnarray}

Substituting Eq. (\ref{LT2}) and (\ref{LT6}) in Eq. (\ref{LT5}), we have
\begin{eqnarray}\label{LT7}
\Delta\alpha_n= - \frac{\bar{p}D_{OL}}{\beta_c{e_n}}\left(\theta-\theta^{0}_n\right).
\end{eqnarray}

Therefore, substituting Eq. (\ref{LT7}) and Eq. (\ref{LT1}), we have
\begin{eqnarray}\label{LT8}
\theta_n \approx \theta^{0}_n + \left(\frac{{e_n}\beta_c}{\bar{p}}\right) \frac{D_{OS}(\psi-\theta^{0}_n)}{D_{OL}D_{LS}}.
\end{eqnarray}

Although the deflection of light preserves the surface brightness, gravitational lensing alters the appearance of the solid angle of the source. Thus, the total flux received by a lensed image is proportional to its magnification $\mu_n$, which is defined by $\mu_n=\Big| \frac{\psi}{\theta}\frac{\partial\psi}{\partial\theta}\mid_{\theta=\theta^{0}_n}\Big|^{-1}$. Therefore, substituting Eq. (\ref{LT1}) into Eq. (\ref{LT7}), we have
\begin{eqnarray}\label{LT9}
\mu_n= \frac{e_n(1+e_n)}{\psi\bar{p}}\left(\frac{\beta_c}{D_{OL}}\right)^2\frac{D_{OS}}{D_{LS}}.
\end{eqnarray}
The amplification decays very quickly with $n$, this indicates that the brightness of the first image $\theta_1$ dominates over the others. On the other hand, the presence of the factor $\left(\frac{\beta_c}{D_{OL}}\right)^2$ indicates that the amplification is always small. It can also be noted that in the limit $\psi\to{0}$ in which maximum alignment occurs between the source, the lens and the observer, amplification should diverge, thus maximizing the possibility of detecting relativistic images.

\subsection{Observables in the strong field limit}\label{sec51}
Finally, we can express the position of the relativistic images, as well as their fluxes, in terms of expansion coefficients ($\bar{p},\bar{b}$, and $\beta_c$). We will now consider the inverse problem, that is, determining the expansion coefficients from observations. In this way, we can understand the nature of the object that generates the gravitational lens and compare the predictions made by modified gravity theories. The impact parameter may be written in terms of $\theta_\infty$ \cite{BOZZA},
\begin{eqnarray}\label{LT10}
\beta_c= D_{OL}\theta_\infty.
\end{eqnarray}

We will follow Bozza \cite{BOZZA} and assume that only the outermost image $\theta_1$ is resolved as a single image while the others are encapsulated in $\theta_\infty$. Thus, Bozza defined the following observables,
\begin{eqnarray}\label{LT11}
s&=&\theta_1-\theta_\infty=\theta_\infty{e^{\frac{\bar{b}-2\pi}{\bar{p}}}}, \\\label{LT12}
\tilde{r} &=& \frac{\mu_1}{\sum^\infty_{n=2}\mu_n}=e^{\frac{2\pi}{\bar{p}}}.
\end{eqnarray}

In the above expressions, $\textbf{s}$ represents the angular separation and $\tilde{r}$ the connection between the flux of the first image and the flux of the rest. These forms can be inverted to obtain the expansion coefficients. To analyze the observables, let us consider that the object in question has an estimated mass of $4,4\times{10^6}M_{\odot}$ and is at an approximate distance of $D_{OL}=8,5Kpc$, these data are the same for the BH at the center of our galaxy \cite{BHCENTER}. Knowing that the critical impact parameter, $\beta_c=3\sqrt{3}M$, does not depend on the radius of the throat, $\textbf{a}$, we then obtain it directly. In geometric units, we have the rescaling of the mass to $M\to \frac{MG}{c^2}$ and $\theta_\infty=26.5473\mu{arcsecs}$, which is the same used for the Schwarzschild BH.

\begin{figure}[htb!]
\centering  
	\subfigure[]{\label{LT1G}
	{\includegraphics[width=0.47\linewidth]{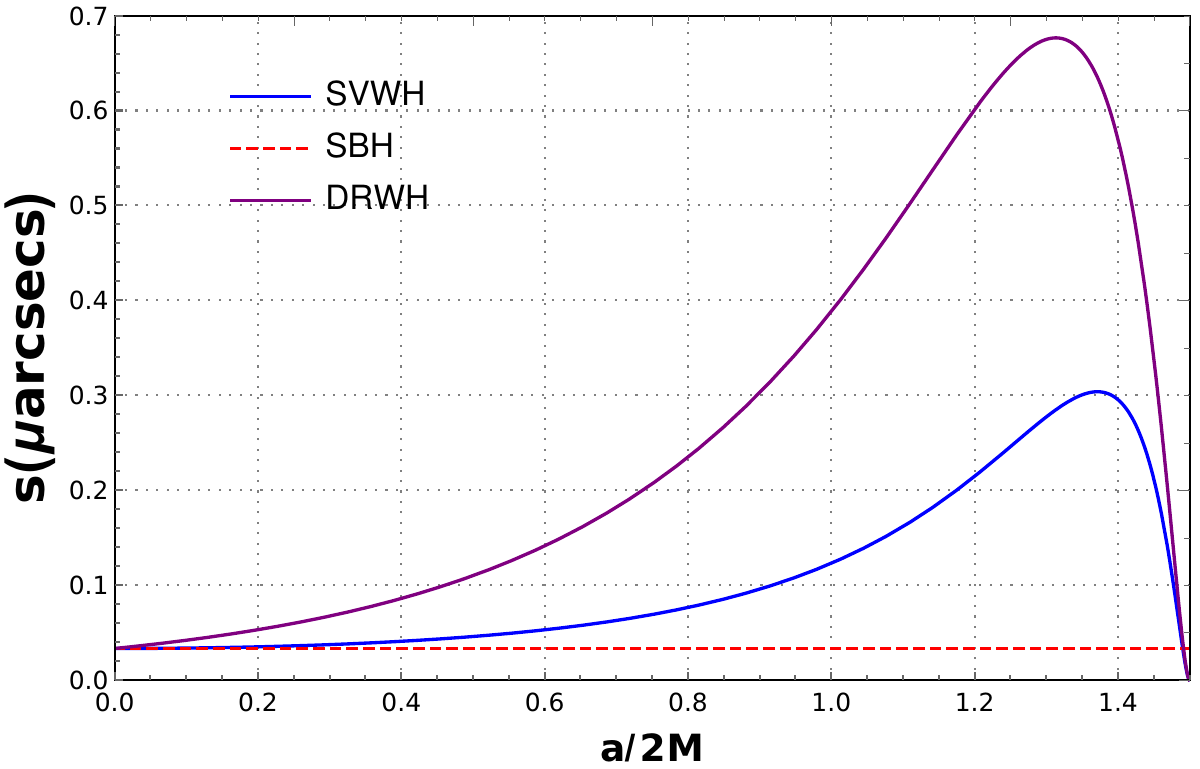}}}\qquad
	\subfigure[]{\label{LT2G}
	{\includegraphics[width=0.47\linewidth]{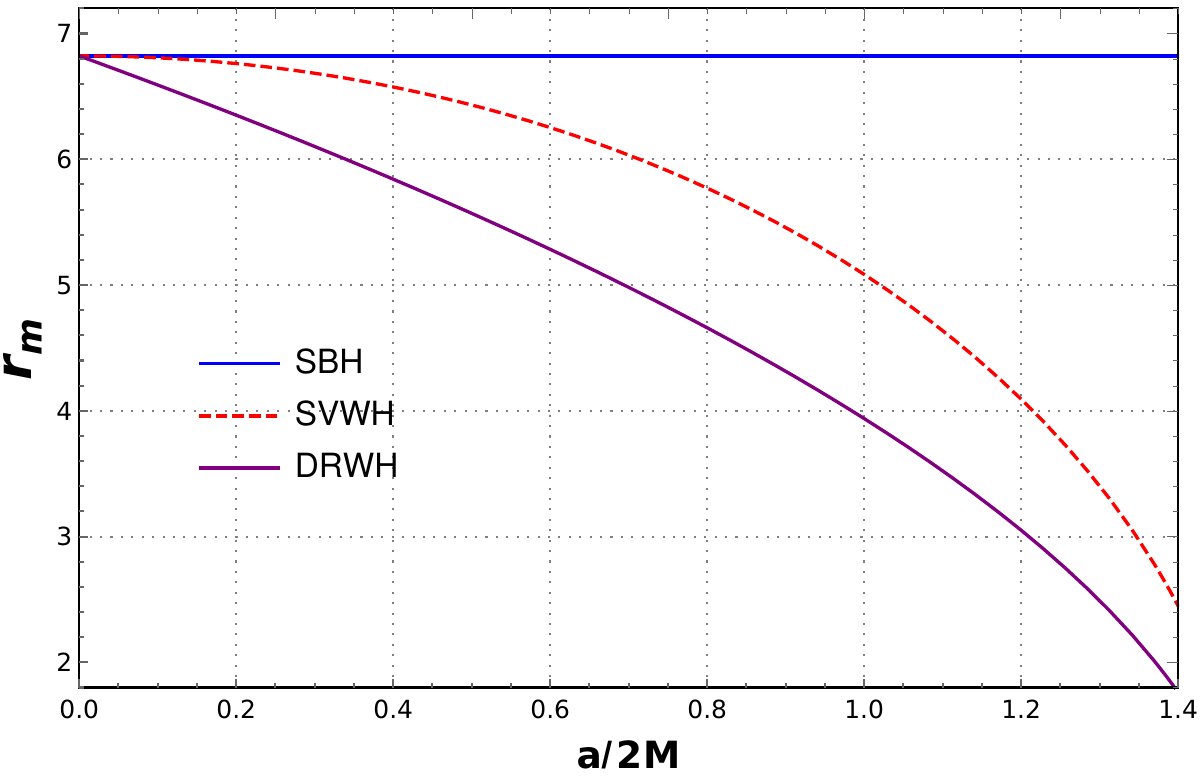}}}
\caption{In (a) we have angular separation $\textbf{s}$ as a function of $a/2M$ and in (b) $2.5\log_{10}\tilde{r}$ as a function of $a/2M$.}
\label{LENTE1}
\end{figure}

In Fig. (\ref{LENTE1}), we have a graphical representation for the angular separation $\textbf{s}$ and $2.5\log_{10}\tilde{r}$ as a function of the ratio between the parameters $a/2M$. In the left panel, we have the curves that represent the angular separation for 3 models, with the blue curve representing the SV case, the purple curve the DR model, and the dotted red curve the Schwarzschild BH, while in the right panel the purple solid line describes the de DR case, the red dotted line represents the SV case, and the Schwarzschild BH is represented in the blue line. In both the SV and the DR models, the angular separation reaches a maximum point and then decreases toward the event horizon, and this behavior can still be seen in the table of observables \ref{TAB1} for the LQG model. The results for SV are in agreement with those analyzed in \cite{SOARES1} and with the Schwarzschild limit.

\begin{table}[ht]

\centering
\caption{observables}

\label{TAB1}
\def\arraystretch{1.9}

\begin{tabular}{@{}lrrrr@{}}
\toprule

\textbf{$a/2M$} & \textbf{$s({\mu}arcsecs)$} & {\textbf{$r_m(magnitudes)$}} \\

\midrule
0          & $0.0332$  \qquad    & $6.8218$      \\
0.2        & $0.0529$   \qquad   & $6.3508$      \\
0.5        & $0.1099$   \qquad   & $5.5700$      \\
0.8        & $0.2348$   \qquad   & $4.6602$      \\
1.2        & $0.6009$   \qquad   & $3.0508$      \\
1.3        & $0.6749$   \qquad   & $2.4910$      \\
1.4        & $0.5700$   \qquad   & $1.7614$      \\
1.45       & $0.3394$   \qquad   & $1.2455$      \\
\bottomrule

\end{tabular}

\end{table}
As shown in the Fig. (\ref{LENTE1}) and table \ref{TAB1}, the angular separation increases with the parameter \( a \). Specifically, for \( \frac{a}{2M} > 0.5 \), it increases by more than an order of magnitude compared to the other cases. This is particularly noteworthy because, from an observational perspective, it simplifies distinguish the outermost relativistic image from the others.  
We also highlight the decrease in the magnitude, $r_m$, which may indicate an increase in the brightness of the other images. Given that the next-generation \textit{Event Horizon Telescope} (ngEHT~\cite{ngetht}) is expected to achieve sufficient sensitivity to resolve relativistic images~\cite{PM}, we anticipate using observational data in the near future to test these theoretical predictions.

\subsection{Observables in the weak field limit}\label{sec52}

The next step is to analyze the observables in the weak field regime, which means that the impact parameter is very large, $\beta\gg{M}$, so the light beam does not form loops. Therefore, the expression Eq. (\ref{DB8}) becomes
\begin{eqnarray}\label{LTW1}
\delta\phi_{LQG} \sim  \frac{2(2M+a)}{\beta}.
\end{eqnarray}

Regarding the equation that relates the angular position of the matter and the image, Eq. (\ref{LT1}), we will consider the perfect alignment between the matter, the compact object, and the observer, which implies $\psi=0$. In this way, we have to
\begin{eqnarray}\label{LTW2}
\theta= \frac{D_{LS}}{D_{OS}}\Delta\alpha_n,
\end{eqnarray} where $\Delta\alpha_n$ is given by Eq. (\ref{LTW1}). Therefore, substituting Eq. (\ref{LTW1}) into Eq. (\ref{LTW2}) we have
\begin{eqnarray}\label{LTW3}
\theta=\theta_E= \sqrt{\frac{D_{LS}(4M+2a)}{D_{OS}D_{OL}}},
\end{eqnarray} where $\theta_E$ is the angular position of the Einstein ring. Therefore, it is clear from the above expression that the presence of the throat radius is modifying the position of the Einstein ring which comes exclusively from LQG effects. To be convinced of this statement, simply compare the expressions (\ref{LTW1}) and (\ref{DB6}). We can also calculate the radius of the Einstein ring from the approximation $\beta\approx\theta{D_{OL}}$ and using Eq. (\ref{LTW3}) \cite{ANEL1,ANEL2}. Therefore, we have
\begin{eqnarray}\label{LTW4}
R_E= D_{OL}\theta_E.
\end{eqnarray} In this way, we will be able to calculate the observables $R_E$ and $\theta_E$ taking into account some values of the ratio between the throat parameter $\textbf{a}$ and mass $\textbf{M}$. We will consider for the procedure of our analysis in this weak field regime the lensing for a bulge star \cite{BOJO}. To this end, we consider the following values for the parameters $D_{OL}=4Kpc$ and $D_{OS}=8Kpc$. In this perspective, in table \ref{TAB2} we consider some values of the ratio between the throat radius and the mass of the BH modified by LQG and calculate the corresponding observables for the previously established scenario and the viable theoretical values of these observables are within the spectrum of possible detection.

\begin{table}[ht]

\centering
\caption{observables}

\label{TAB2}
\def\arraystretch{1.9}

\begin{tabular}{@{}lrrrr@{}}
\toprule

\textbf{$a/2M$} & \textbf{$R_E (km)$} & {\textbf{$\theta_E (arcsecs)$}} \\

\midrule
0          & $1.27\times{10^{12}}$  \qquad    & $2.12$      \\
0.2        & $1.39\times{10^{12}}$   \qquad   & $2.32$      \\
0.4        & $1.50\times{10^{12}}$   \qquad   & $2.50$      \\
0.8        & $1.70\times{10^{12}}$   \qquad   & $2.84$      \\
1.2        & $1.88\times{10^{12}}$   \qquad   & $3.14$      \\
1.4        & $1.96\times{10^{12}}$   \qquad   & $3.28$      \\
\bottomrule

\end{tabular}

\end{table}

\section{CONCLUSION}\label{sec6}

In the present work, we theoretically investigate gravitational lensing effects for two spacetimes that represent BH solutions recently proposed as corrections to LQG. For both models investigated, we found analytical expressions for the deflection of light in the weak field regime Eqs. (\ref{14}) and (\ref{DB6}). In the expression for the deflection of light subjected to the effects of LQG of the second model Eq. (\ref{DB6}) we recovered the results related to SV spacetime \cite{SOARES1} when $\bar{a}\to{0}$. We performed the same analysis for the light deflection in the strong field regime and obtained analytical expressions for the contribution of the divergent part of the integration Eqs. (\ref{23}) and (\ref{DB16}). Due to technical difficulties, we needed to perform a numerical analysis of the regular part of the integration and consequently of the total angular deviation. The numerical analysis for the angular deflection was performed for both models as a function of the impact parameter \textbf{${\beta}$} for some values of the bounce parameter \textbf{a}.

In the second part of our analysis of the second model, we use the calculation of the angular deflection of light in the gravitational lensing equations to relate the observables to relativistic images, both in the strong field and weak field regimes. For the strong field regime, we model our calculation of angular deflection with data for a BH at the center of our galaxy \cite{BHCENTER}. We constructed a representation for the angular separation and verified that it increases according to the effects of LQG and the relativistic image increases in intensity. We can also recover the same results from \cite{SOARES1} when we turn off LQG.

Finally, in the weak field regime for the second model, we consider the data related to a bulge star \cite{BOJO} and analyze the effects of the LQG correction in the calculation of the Einstein ring and its angular position for some values between the ratio of the WH throat parameter and the mass. Naturally, the values of these observables differ from those predicted by the  Schwarzschild solution due to the LQG effects and we hope that due to both the analytical and numerical results developed throughout this work it can contribute to the verification of models that go beyond general relativity. As a future perspective, we know that regular spacetimes have as matter content some class of nonlinear electrodynamics or a combination of nonlinear electrodynamics with a phantom scalar field. So we can try to reconstruct their matter Lagrangian with the presence of LQG correction terms and then analyze whether or not the violation of the energy conditions of the system occurs. It is also important to note that astrophysical black holes are not static and must possess rotation. Therefore, in future work, we should investigate how rotation influences the capacity of these solutions to reproduce observable data.

\begin{acknowledgments}
M. S. would like to thank Funda\c c\~ao Cearense de Apoio ao Desenvolvimento Cient\'ifico e Tecnol\'ogico (FUNCAP) for the financial support. H. Belich would like to thank CNPq for financial support.
\end{acknowledgments}

\clearpage

\nocite{*}
		
\end{document}